\documentclass[pdflatex,sn-mathphys-num]{sn-jnl}

\usepackage{graphicx}%
\usepackage{multirow}%
\usepackage{amsmath,amssymb,amsfonts}%
\usepackage{amsthm}%
\usepackage{mathrsfs}%
\usepackage[title]{appendix}%
\usepackage{xcolor}%
\usepackage{textcomp}%
\usepackage{manyfoot}%
\usepackage{booktabs}%
\usepackage{algorithm}%
\usepackage{algorithmicx}%
\usepackage{algpseudocode}%
\usepackage{listings}%
\usepackage{bbm}%
\usepackage{comment}
\usepackage{subcaption}%
\usepackage{cleveref}%
\usepackage{tikz,tikz-3dplot}%

\theoremstyle{thmstyleone}%
\theoremstyle{thmstyletwo}%

\theoremstyle{thmstylethree}%

\begin{document}

\title[Article Title]{A Spatio-temporal CP decomposition analysis of New England region in the US}


\author*{\fnm{Fatoumata} \sur{Sanogo}}\email{fsanogo@bates.edu}



\affil*{\orgdiv{Mathematics Department}, \orgname{Bates College}, \orgaddress{\street{2rd Andrews Road}, \city{Lewiston}, \postcode{04240}, \state{Maine}, \country{USA}}}




\abstract{Spatio temporal data consist of measurement for one or more raster fields such as weather, traffic volume, crime rate, or disease incidents. Advances in modern technology have increased the number of available information for this type of data hence the rise of multidimensional data. In this paper we take advantage of the multidimensional structure of the data but also its temporal and spatial structure. We will be using the NCAR Climate Data Gateway website which provides data discovery and access services for global and regional climate model data. The daily values of total precipitation (prec), maximum (tmax), and minimum (tmin) temperature are combined to create a multidimensional data called tensor (a multidimensional array). In this paper, we propose a spatio temporal principal component analysis to initialize CP decomposition component. We take full advantage of the spatial and temporal structure of the data in the initialization step for cp component analysis. The performance of our method is tested via comparison with most popular initialization methods. We also run a clustering analysis to further show the performance of our analysis.}




\maketitle

\section{Introduction}\label{intro}

Many places have experienced changes in rainfall resulting in floods, droughts, or intense rain (see \cite{climate}) and heat waves. This has caused a lot of concerns around the world and increased the need for solutions such as developing more reliable predictive and diagnostic tools for weather and climate related events. Many natural phenomena such as meteorological or flow variables may be simulated by numerically solving modeling equations. Numerical Earth System Models (ESMs), whether global or regional, remain the principal instruments for simulating atmospheric and hydrological processes by solving the governing physical equations over discretized spatial domains. However, these models face two main challenges: heavy computation times that limit scenario exploration and near-real-time forecasting, which constrains real-time applications, and the exponential growth of data dimensionality as spatial resolution increases. These constraints motivated the development of statistical and machine-learning methods that emulate the physics-based models. Two common tasks have emerged: \textbf{downscaling}, which aims to generate high-resolution fields from low-resolution inputs, and \textbf{forecasting}, which predicts future states at fixed resolution. Convolutional Neural Networks are effective for learning spatial features in downscaling applications \citep{Goodfellow-et-al-2016,WangEtal2021,HohleinEtal2020,AnnauEtal2023}, but their reliance on Cartesian grids complicates their use with the non-rectangular meshes native to most ESMs \citep{article}, making regridding part of pre-processing~\citep{AnnauEtal2023}. Regridding such data can introduce artifacts and blur physically homogeneous regions, potentially obscuring large-scale climate events such as the 2003 European drought \citep{euro}.
Graph neural networks \citep{lam2022graphcast,chen2023fengwu} partially address this issue by operating on irregular grids for forecasting, yet they often demand extensive training resources and high-performance computing infrastructures. 

Dimensionality-reduction methods offer an alternative pathway. Principal Component Analysis (PCA) \citep{Ye_2025,jolliffe2016principal,bishop2006pattern} also known in the climate sciences as Empirical Orthogonal Function analysis \citep{HannachiEtal2007,MonahanEtal2009}—is the standard tool for deriving compact spatial and temporal representations. While PCA efficiently extracts dominant variance patterns, it treats variables independently and imposes orthogonality constraints that lack clear physical interpretation \citep{HannachiEtal2007}.

Tensor decomposition techniques such as Canonical Polyadic decomposition (CPD) are designed to address these limitations. The Canonical Polyadic (CP) decomposition \citep{Kol,KRUSKAL197795,Hitchcock1927TheEO,Carroll,Harsh}, generalize PCA by modeling multi-dimensional relationships across space, time, and variables without enforcing orthogonality. CP expresses a tensor as a sum of rank-one components, enabling interpretable latent-factor discovery and providing a scalable framework for analyzing multiway climate datasets \citep{Aggarwal2015, Lathauwer1996FromMT}. By decomposing a tensor into rank-one components, CP decomposition enables latent factor and hidden patterns or structures discovery in multiple-axes data analysis across many fields. It was introduced by \citet{Hitchcock1927TheEO} as the polyadic decomposition and later by \citet{Carroll} (CANDECOMP) and \citet{Harsh} (PARAFAC); the unified term is CP Decomposition. Although tensor approaches are gaining traction in spatio-temporal data analysis \citep{Kol,Xu,tian2023tensor}, their application to climate research remains limited, especially regarding how spatial and temporal dependencies are incorporated into the initialization and estimation process. These factorizations yield low-dimensional, mode-specific factors (e.g., space, time, variable) and are often used as the basis for a clustering analysis to reveal coherent climate regimes. 
In fact, climate and meteorological studies often rely on categorizing data (observations or variables) into distinct subgroups with similar characteristics within groups (i.e. clustering). This is essential, as it allows for climate regionalization, it identify patterns such synoptic types \cite{article12}. Clustering is a powerful tool that can help facilitate the analysis of meteorological events. In forecast analysis and evaluation, clustering can help in assessing the uncertainty and performance of forecasting models. It is often preceded by a dimensionality reduction technique like PCA due to the high dimensionality nature of climate data \cite{article09, article12}.

In this work, we propose a spatio-temporal initialization strategy for CP decomposition that explicitly exploits the inherent structure of climate data. By initializing factor matrices using spatio-temporal principal components, our approach improves factor identifiability, and reconstruction accuracy, compared with standard methods such as random or HOSVD-based initialization. Furthermore, we applied clustering to the resulting factor matrices, with performance assessed via silhouette scores.
Our main contributions are:

(1) A novel initialization procedure for CP decomposition that leverages spatio-temporal correlations to enhance reconstruction quality.

(2) An evaluation on NCAR precipitation and temperature datasets, against common initialization methods.

(3) A clustering analysis illustrating the performance of our model.
\\~\\
The remainder of the paper is organized as follows. Section \ref{regdata} describes the studied dataset and its spatial–temporal structure. Section \ref{spatiotemp:decomp} details the proposed methodology. Section \ref{result} presents experimental results and comparative analyses.

\section{Data overview}\label{regdata}

For our analysis we considered a dataset, which includes three variables total precipitation (prec), maximum temperature (tmax) and minimum temperature (tmin) daily value from January 1st, 1979 to December 31st, 2024. These variables are important for hydrological type studies. The variables were sourced from the NA-CORDEX data collection, available via the NCAR Climate Data Gateway website (\url{https://doi.org/10.5065/D6SJ1JCH}). They are output from the Canadian Regional Climate Model (CRCM5-OUR), which was run under the evaluation scenario forced by ERA-Interim reanalyses. To focus on information solely provided by the regional climate model we use the data with no interpolation or bias correction. The database covers North America with a regular grid (native rotated-pole grid) with 340 cells in longitude and 300 cells in latitude. The spatial resolution is 0.22\textdegree (50~km)~\citep{nacordex}. For this paper, we restricted the domain to the New England - which includes six states: Maine, New Hampshire, Vermont, Massachusetts, Rhode Island, and Connecticut -  region in the United States, 31 cells in longitude and 34 cells in latitude for a total of 1054 grid cells.

\subsection{Notations}

Let $1 \leq i \leq I$ a time index, $1 \leq j \leq J$ a location index and $1 \leq k \leq K$ an index that represents the physical variable of interest. Let $\boldsymbol{x}_{i,k} \in \mathbbm{R}^J$ be a vector representing all the values of variable $k$ at time step $i$ over all the locations. Let $\boldsymbol{X}_k \in \mathbbm{R}^{I\times J}$ be a matrix that combines $\boldsymbol{x}_{i,k}$ row-wise and contains all the values of variable $k$ at all time steps and all locations. Lastly, let $\mathcal{X} \in \mathbbm{R}^{I\times J \times K}$ be a mode 3 tensor that stacks $\boldsymbol{X}_k$ for all the variables $k.$ 

In our case, we consider $\boldsymbol{X}_1$, $\boldsymbol{X}_2$ and $\boldsymbol{X}_3 \in \mathbbm{R}^{13149 \times 1054}$ be the matrices holding, respectively, the prec, tmax and tmin data described above. Moreover, let $\mathcal{X} \in \mathbbm{R}^{13149 \times 1054 \times 3}$ be the mode 3 tensor stacking all three climate variables together.

\subsection{Spatial properties}

\begin{figure}[H]
   \centering
    \includegraphics[width=4.29cm, height=3.4cm ]{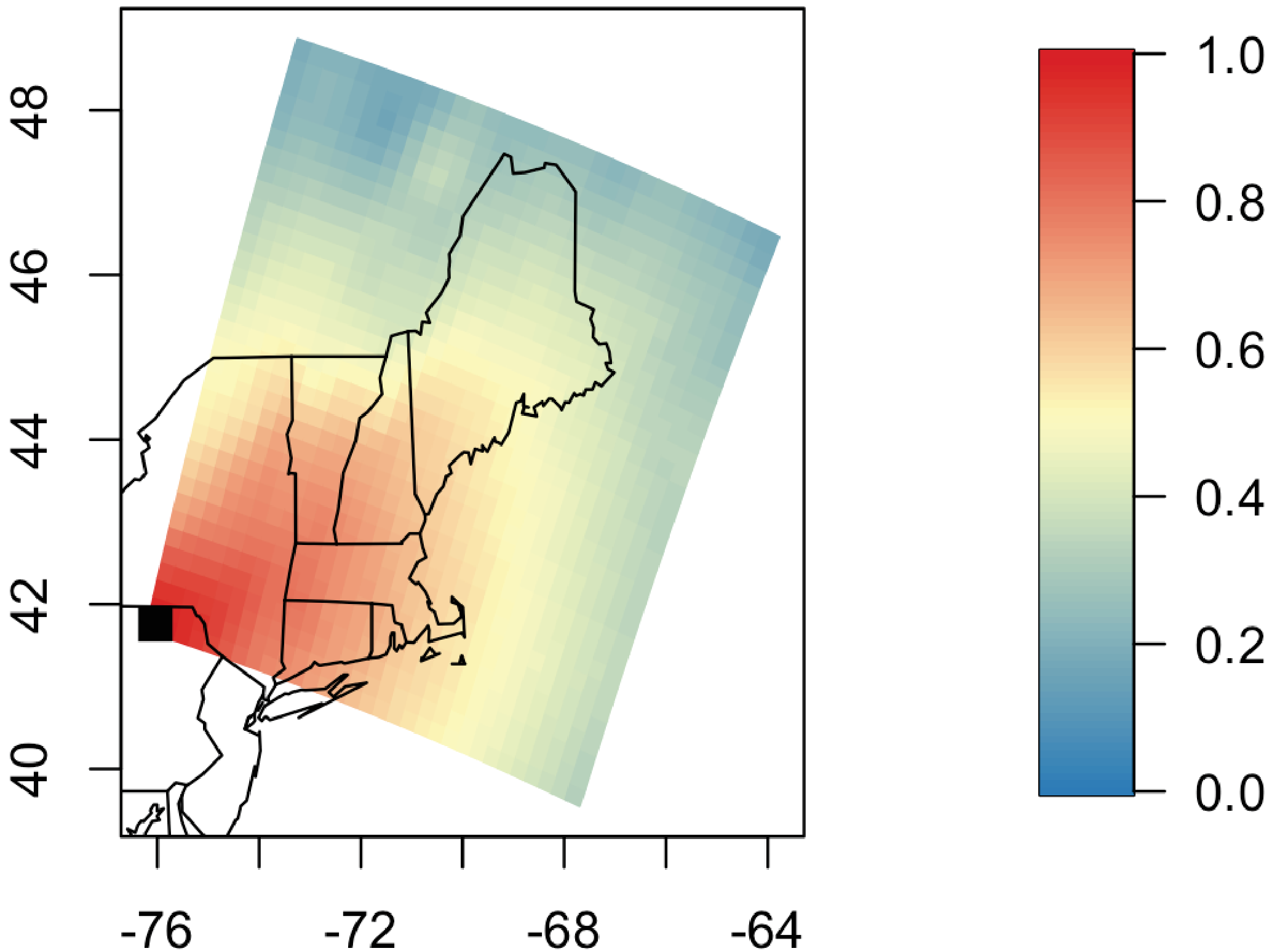}
    \includegraphics[width=4.29cm, height=3.4cm]{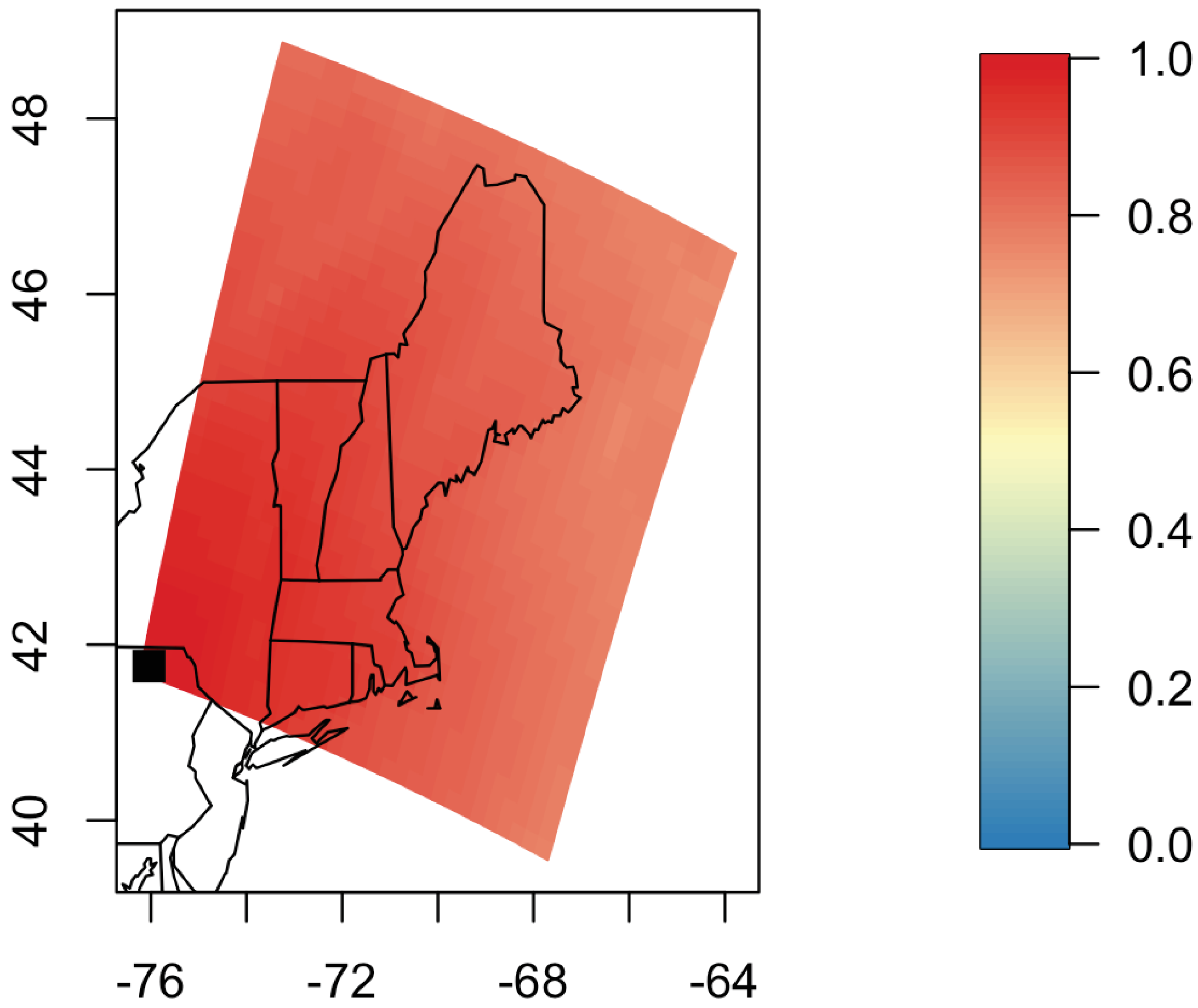}
    \includegraphics[width=4.29cm, height=3.4cm]{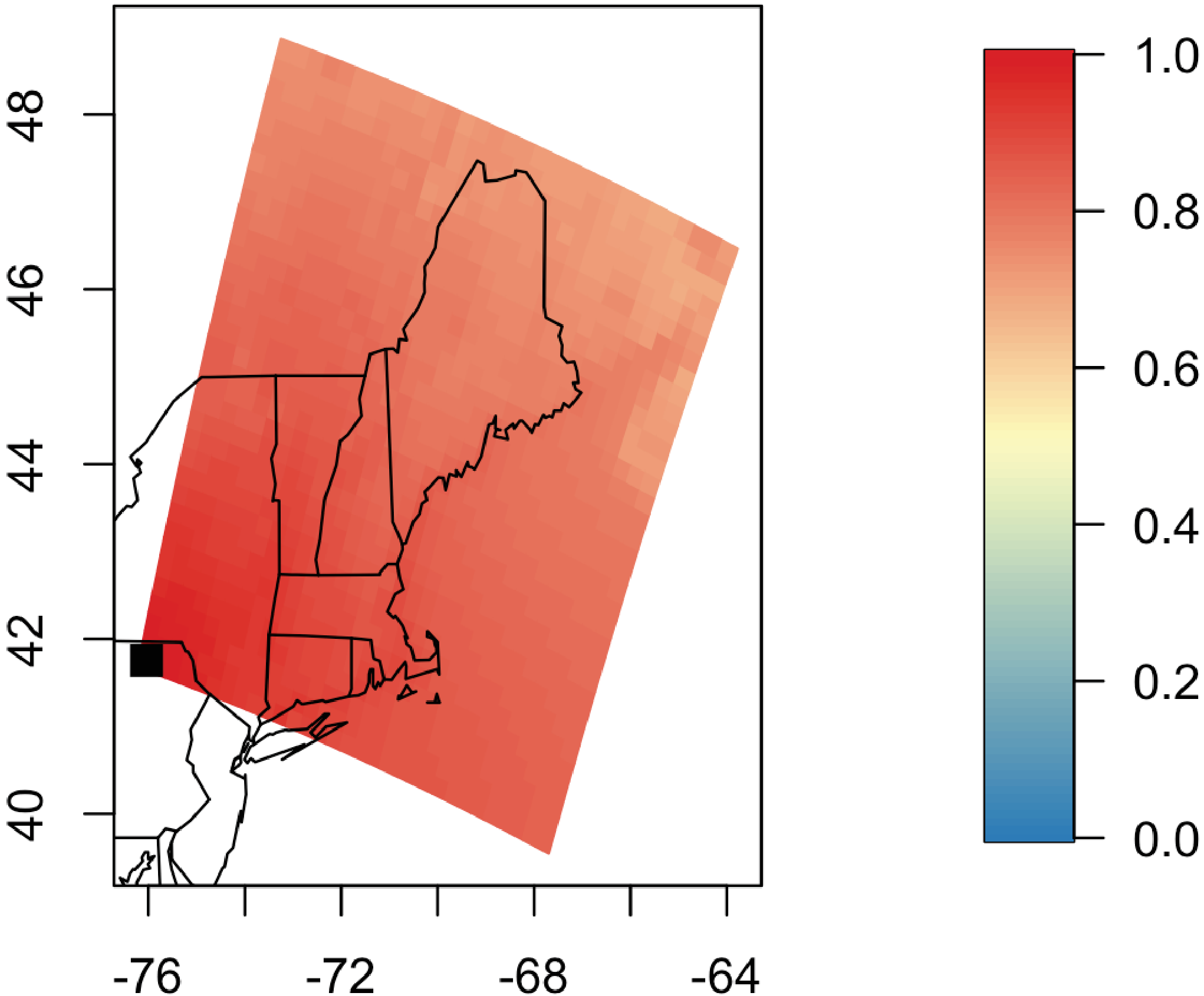}

    \includegraphics[width=4.29cm, height=3.4cm]{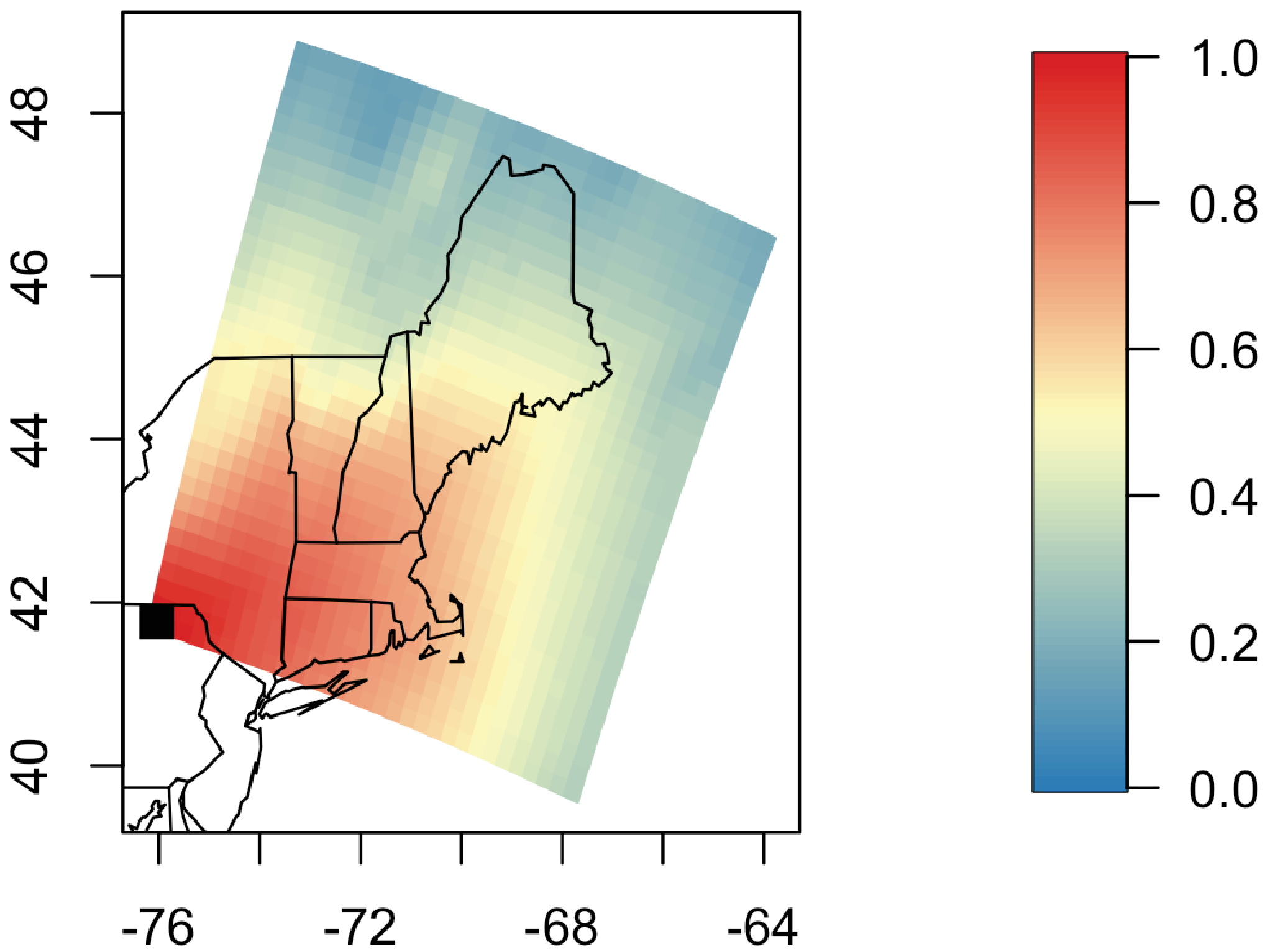}
    \includegraphics[width=4.29cm, height=3.4cm]{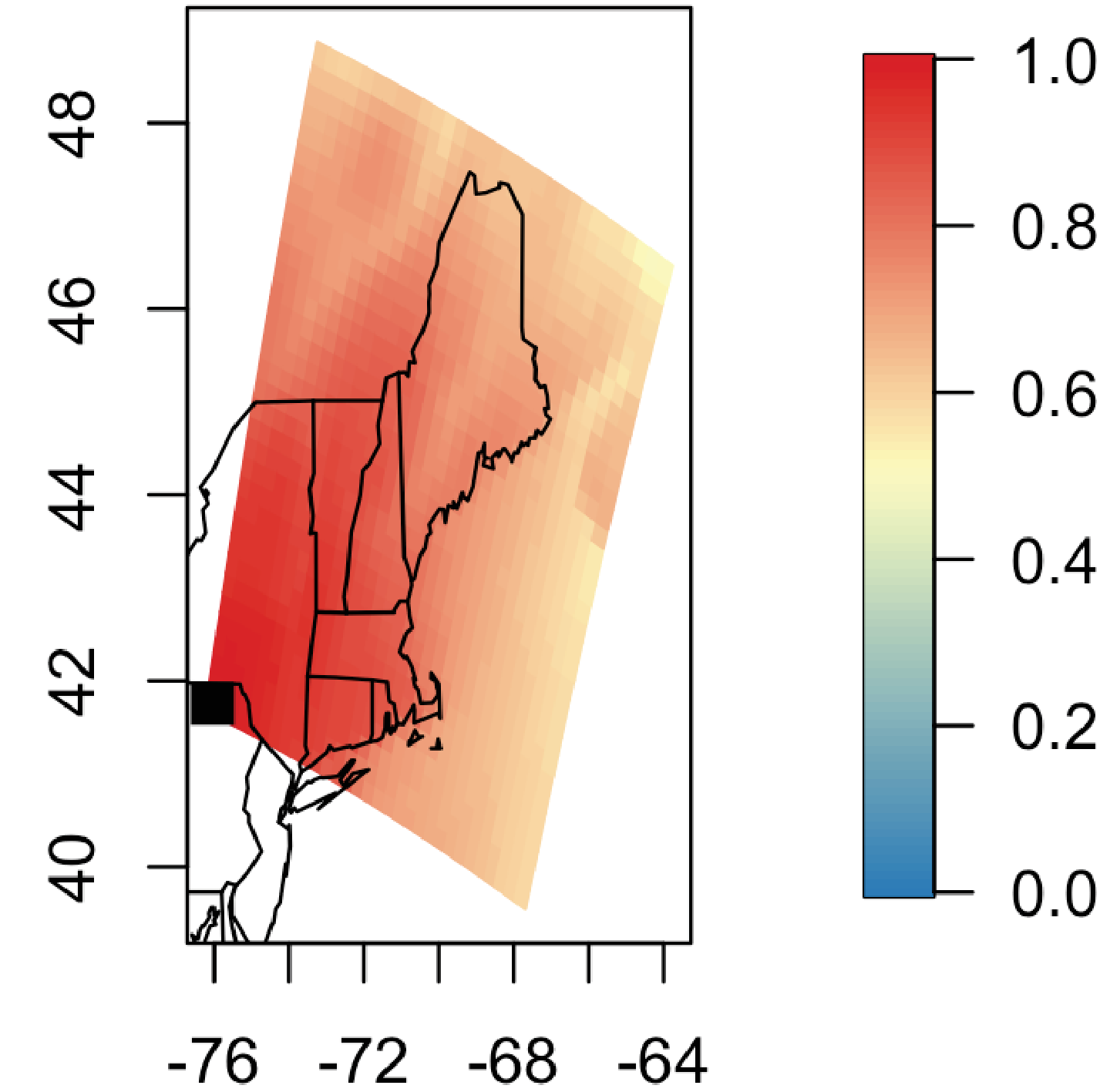}
    \includegraphics[width=4.29cm, height=3.4cm]{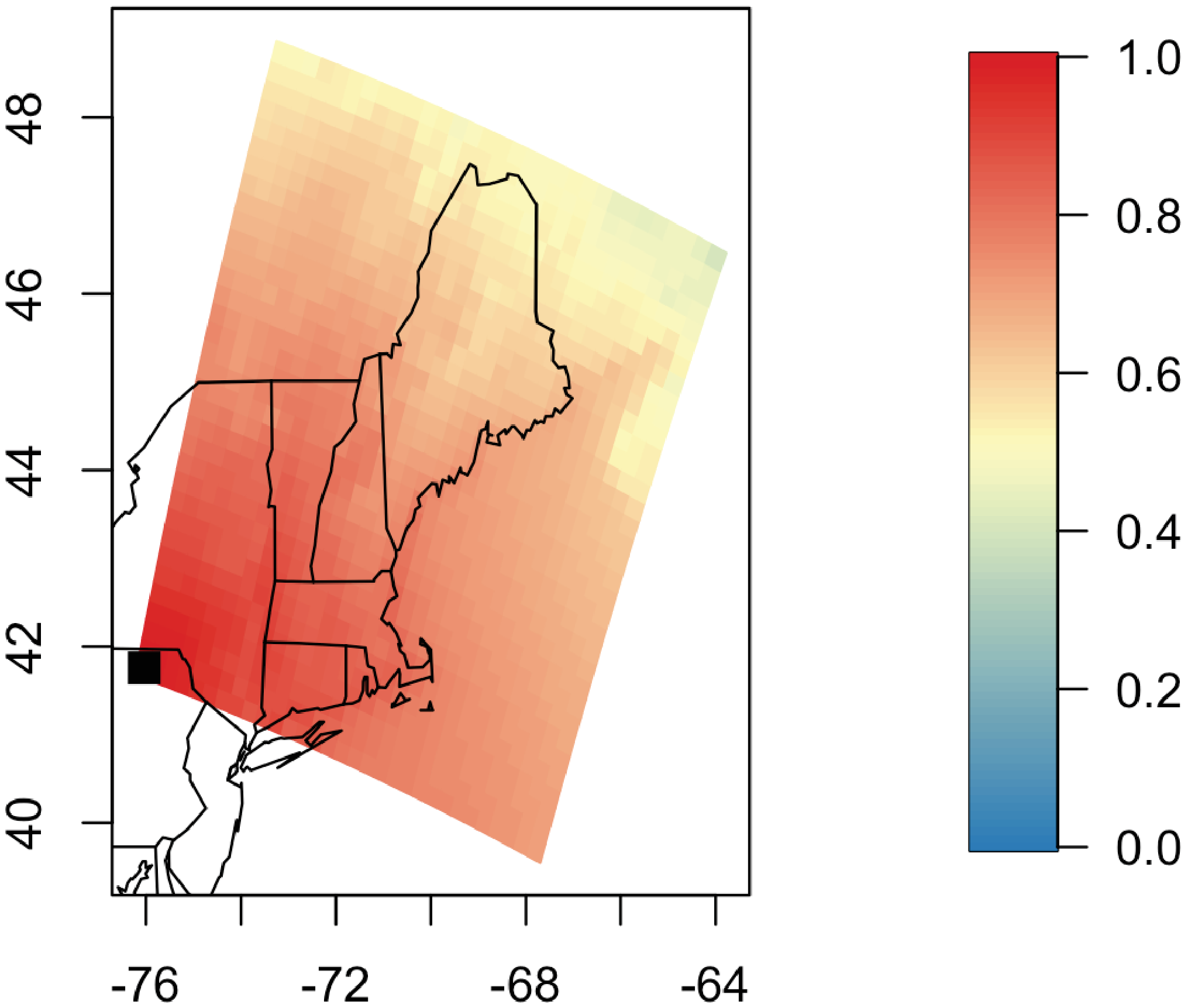}

    \includegraphics[width=4.29cm, height=3.4cm]{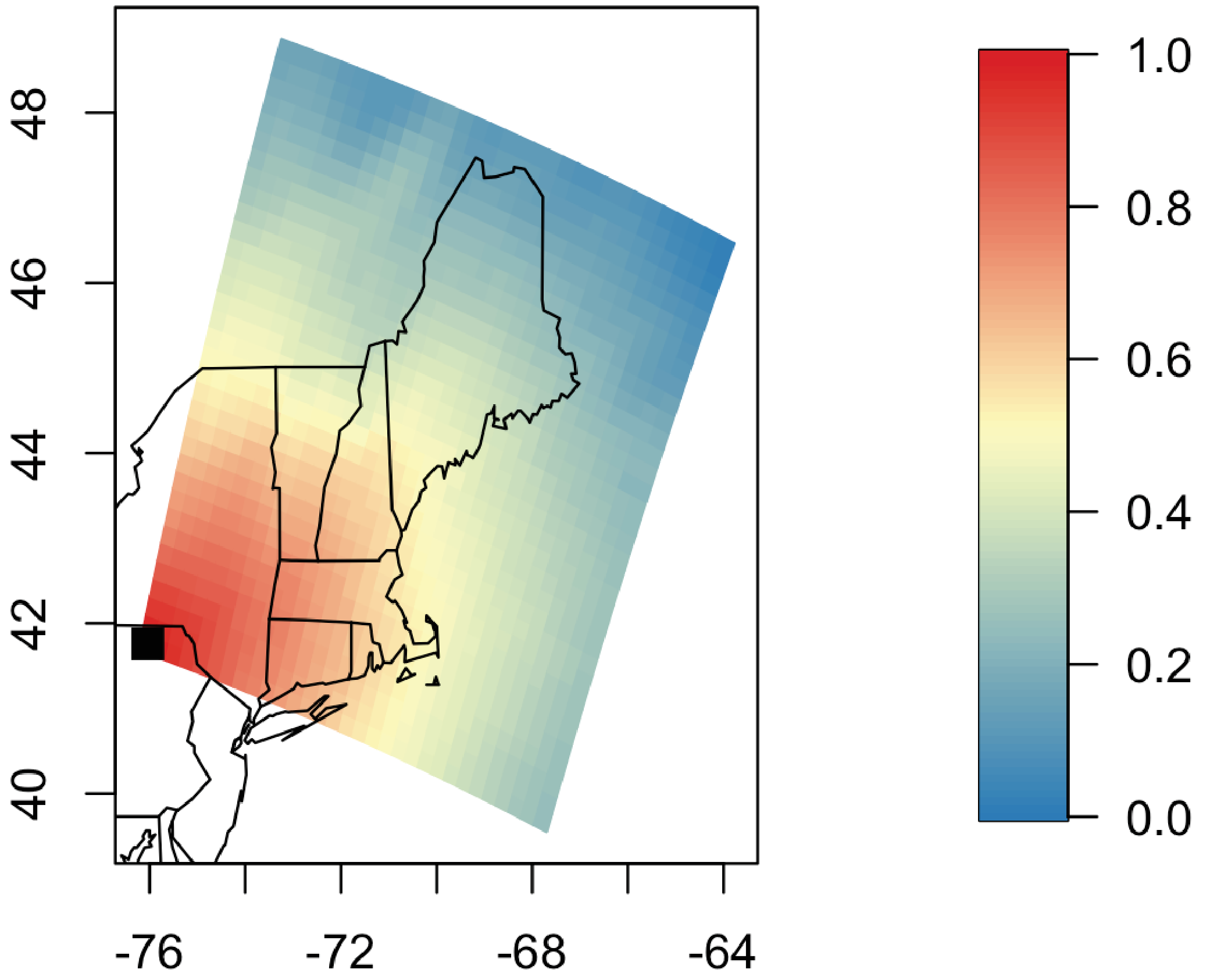}
    \includegraphics[width=4.29cm, height=3.4cm]{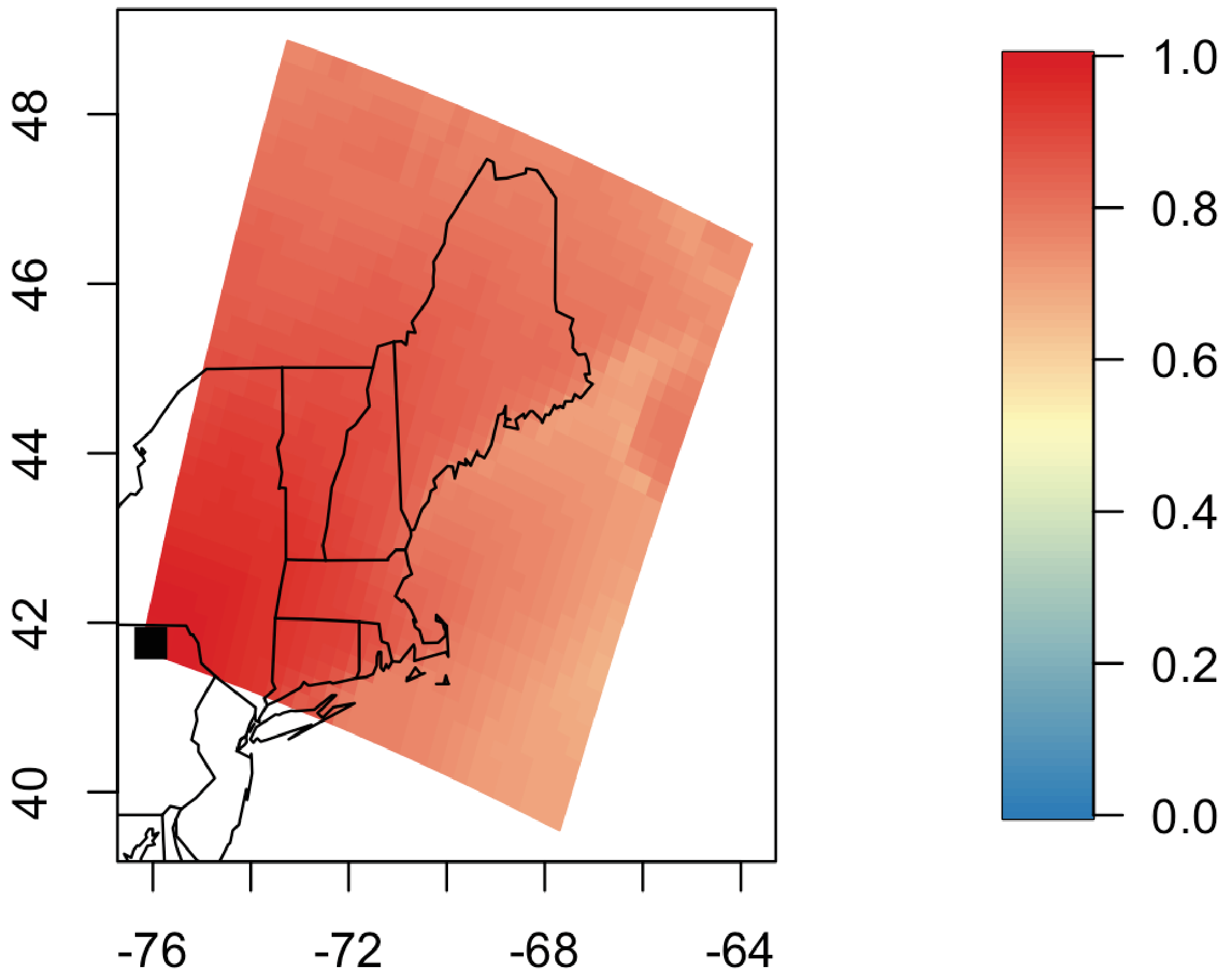}
    \includegraphics[width=4.29cm, height=3.4cm]{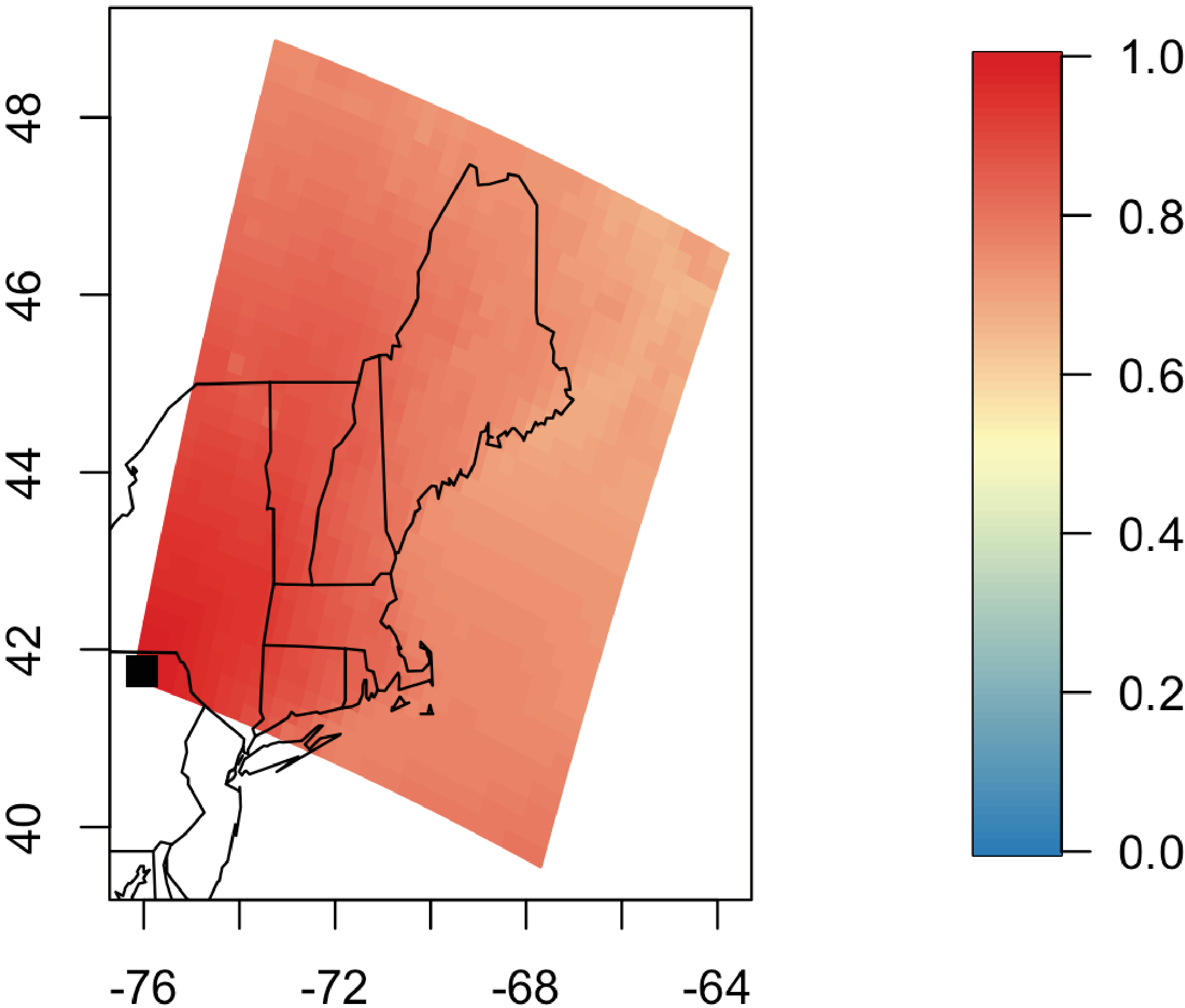}

    \includegraphics[width=4.29cm, height=3.4cm]{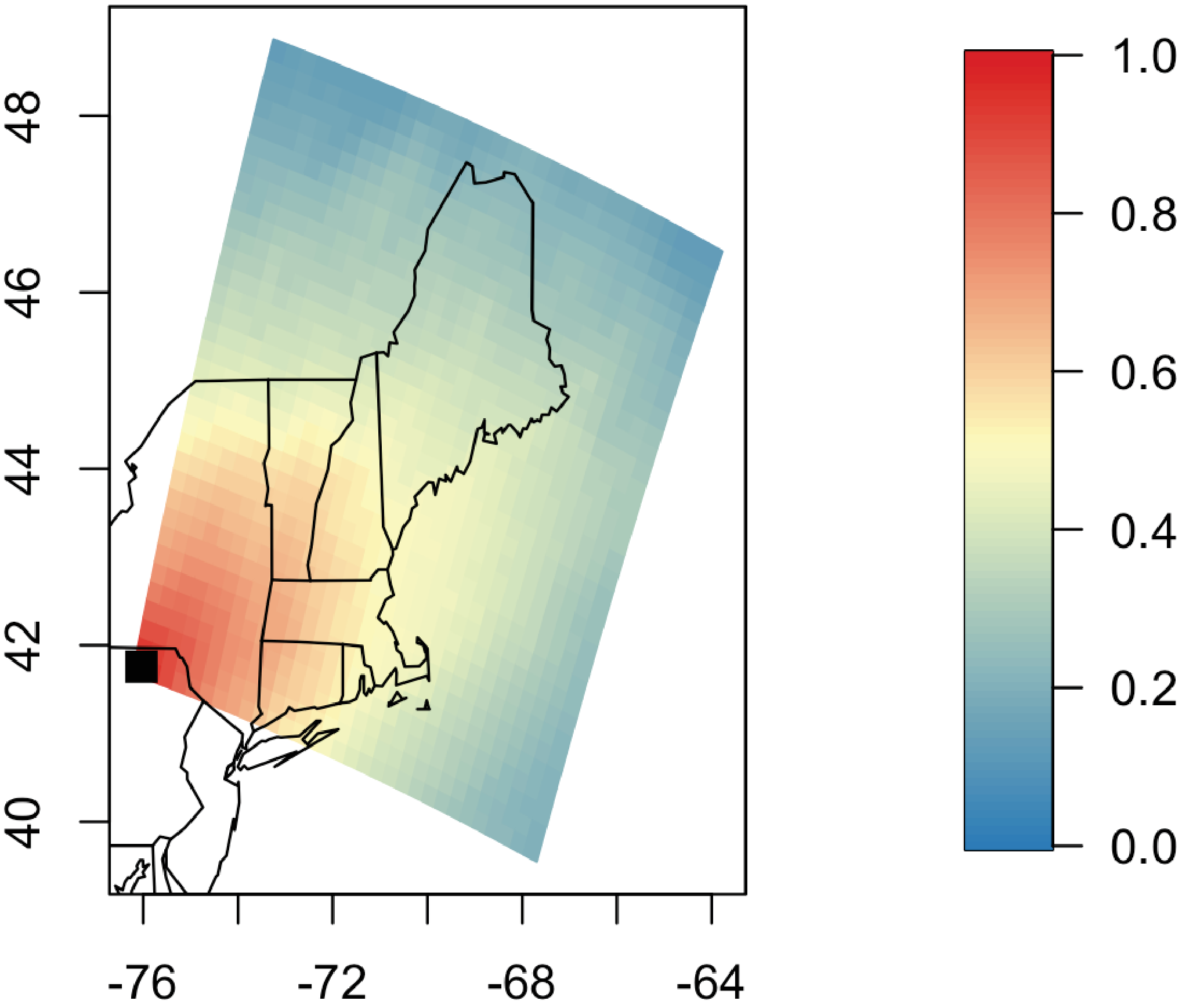}
    \includegraphics[width=4.29cm, height=3.4cm]{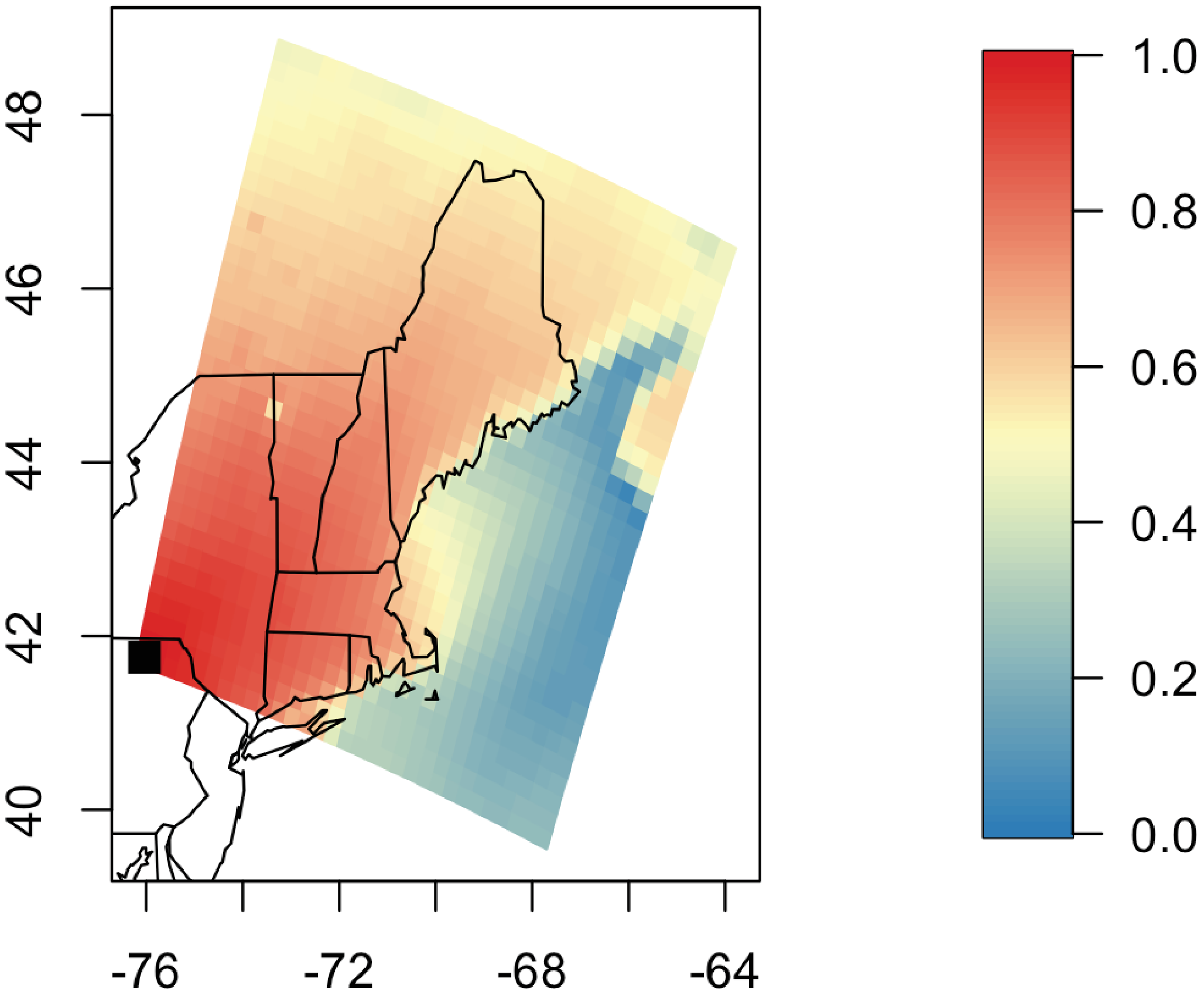}
    \includegraphics[width=4.29cm, height=3.4cm]{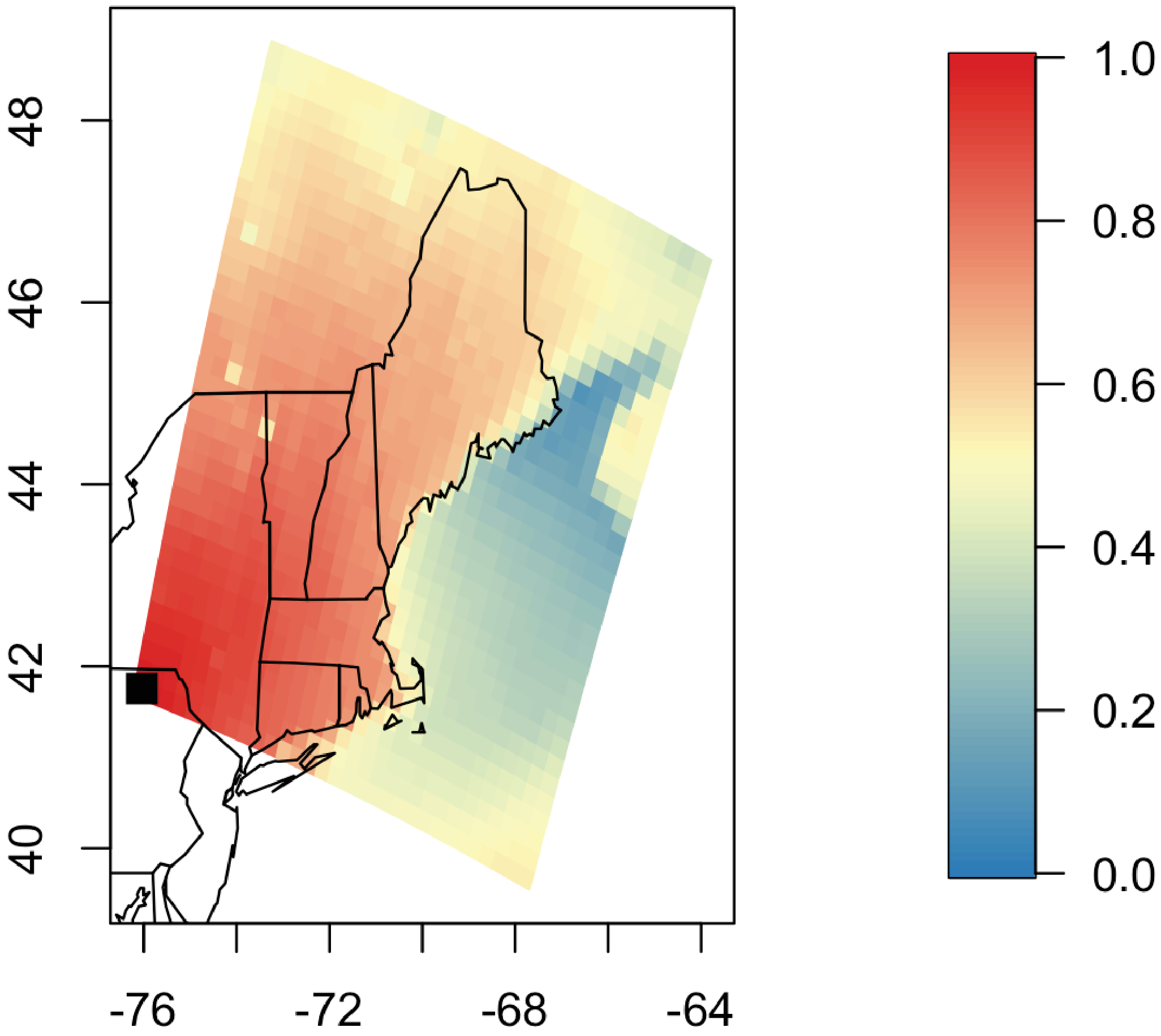}

\caption{Spearman correlation coefficients in fall (Sept-Nov), winter (Dec-Jan), spring (Mar-May), summer (Jun-Aug) (in rows) with respect to the location indicated by the black square tile for prec, tmax and tmin (in columns).}
    \label{fig:exp:ana:corcoef}
\end{figure}

Here our analysis focus on the spatial properties of the climate variables for each of the four continental seasons: winter (from December to January), spring (from March to May), summer (from June to August) and fall (from September to November). To emphasize spatial patterns, we look at maps of Spearman correlation coefficients. Figure~\ref{fig:exp:ana:corcoef} shows Spearman correlation coefficients computed with respect to the reference location indicated by the black square (same color scale across all panels). In all panels, correlations decrease with distance from the reference location. The patterns for tmax and tmin are broadly similar, with slight differences in winter and summer compared to precipitation's. tmax and tmin correlations remain strong in all seasons, with minor reductions in spatial extent during summer. Precipitation (left column) exhibits weaker and more spatially variable correlations compared to temperature. The correlation are all positive.

\subsection{Temporal properties}

\begin{figure}[ht]
    \centering
\includegraphics[width = 13cm]{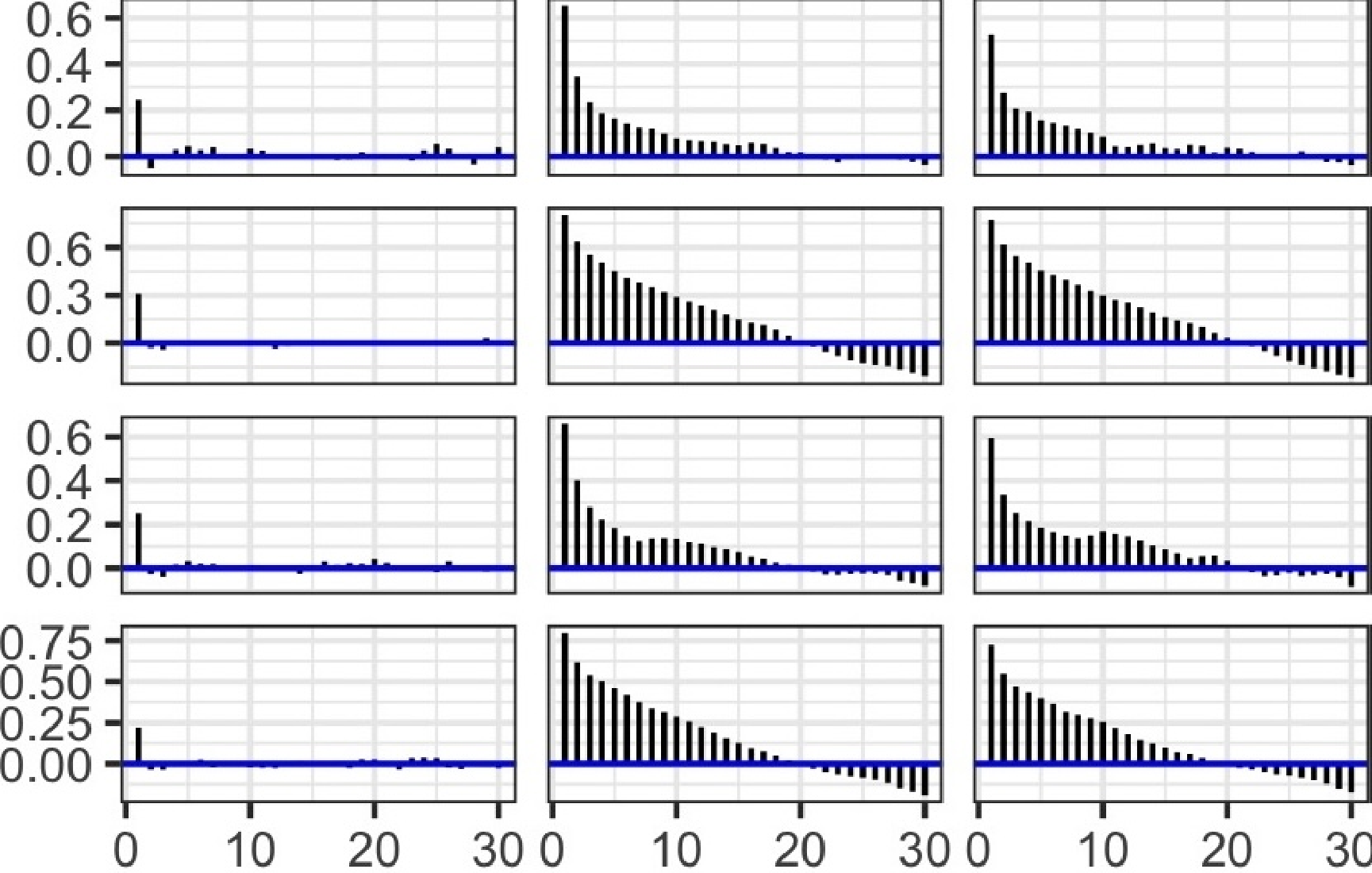}
\caption{Winter (Dec-Jan), spring (Mar-May), summer (Jun-Aug) and fall (Sept-Nov) auto-correlation functions with lags expressed in days (in rows) for prec, tmax and tmin (in columns) at grid cell whose linear index is halfway through the flattened list of cells.}
\label{fig:exp:ana:acf}
\end{figure}

Next, we focus on the temporal properties of the climate variables by looking at the auto-correlation functions per season at a given grid box, see Fig.~\ref{fig:exp:ana:acf}. Again we use Spearman correlation coefficients. The conclusions are similar to those drawn from the spatial patterns in Fig.~\ref{fig:exp:ana:corcoef}. As seen in \ref{fig:exp:ana:acf} the auto-correlation function of prec (first column) is very weak, much weaker than in the spatial case. The temporal dependence of tmax and tmin on the other hand shows clear variation with the season, with stronger dependence in spring and fall. Moreover, the shapes of the auto-correlation functions of tmax and tmin are very much alike.

\section{Methodology} \label{spatiotemp:decomp}

\subsection{Spatio-Temporal Principal Component Analysis}\label{stpca}

Principal component analysis (PCA) is a powerful and widely used method in multivariate statistics. Principal Component Analysis (PCA) \cite{hotelling1933analysis} can be performed on the sample covariance matrix \( \mathbf{S} = \tfrac{1}{i}\mathbf{X_k}^\top \mathbf{X_k} \), where \(\mathbf{X_k}\) is an $  I\times J$ centered data matrix (observations of $j$ locations on $i$ time steps). However, this approach do not take into account the dependence among nearby locations, when applied to spatial (geographic) data. The eigenvalues are nonnegative, and the principal components are given by the eigenvectors associated with the positive eigenvalues.  \citet{cliff1ord} addressed this by introducing the concept of `spatial autocorrelation'. Incorporating spatial autocorrelation \citet{chen2023spatial} into PCA requires specifying a spatial-weights matrix $\mathbf{W}$ that encodes neighborhood or proximity relations among the spatial units under study. Several literature on spatial PCA (SPCA) discusses how to define $\mathbf{W}$ \citep{Cartone03042021,7846858}.
For this paper, our focus will be on Spatio Temporal Principal Component Analysis (STPCA); see \cite{Krzyko02012024}. STPCA treats feature observations as functional time series, emphasizing their temporal evolution while preserving spatial structure. Its aim is to extract the dominant modes of variability in spatio-temporal data by jointly incorporating temporal dynamics, spatial relationships, and variation in feature magnitudes. To run an STPCA, a list of matrices is considered with the rows representing $i$ time steps and columns $j$ locations or region of values of the $k$ features/variables being considered. In fact, first the data matrix is transformed into a multivariate functional data set; see~\cite{gorecki2018selected,gorecki2012functional}. The transformation into a functional data is done by creating a matrix $A$ of size $I\times pk$ through fourier basis \cref{eq:fourier-basis} and keeping the coefficients. \begin{equation}
\label{eq:fourier-basis}
\begin{aligned}
\phi_{0}(t)      &= \frac{1}{\sqrt{n}},\\
\phi_{2b-1}(t)   &= \sqrt{\frac{2}{n}}\;\sin\!\left(\frac{2\pi b\, t}{n}\right),\\
\phi_{2b}(t)     &= \sqrt{\frac{2}{n}}\;\cos\!\left(\frac{2\pi b\, t}{n}\right),
\end{aligned}
\qquad
\begin{aligned}
&b = 1,2,\ldots,\frac{B_k}{2},\\
&k = 1,2,\ldots,m,\\
&t \in [0,n],
\end{aligned}
\quad \text{with } B_k \text{ even.}
\end{equation}
The matrix $A$ is then link to the spatial weight $\mathbf{W}$ using the Moran’s index~\citep{moran1950test} to obtain a spatial correlation matrix. This matrix is then used to find eigenvalues which, unlike standard PCA, can be either positive or negative. The associated set of eigenvectors denotes the spatiotemporal principal components, which are entirely dependent on the choice of $\mathbf{W}$. Changing the matrix $\mathbf{W}$ changes the principal components, and thus the shape of the clusters of objects examined.

\subsection{CP decomposition} \label{CPD}
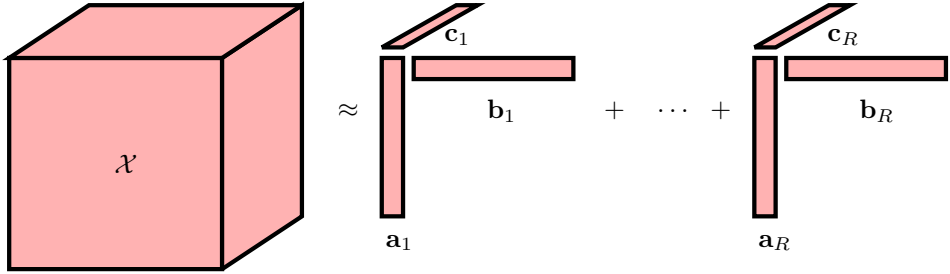
\begin{figure}[ht]
\centering
\begin{tikzpicture}[scale=0.7]
\draw [fill=red!30!,ultra thick] (-2,-2) rectangle (2,2);
\draw[fill=red!30!,ultra thick] (2,-2) -- (3.5,-1) -- (3.5,3) -- (2,2);
\draw[fill=red!30!,ultra thick] (3.5,3) -- (-0.5,3) -- (-2,2) -- (2,2);
\draw[ultra thick] (2,-2) -- (2,2) -- (3.5,3);
\filldraw[black] (-0.2,0)   node[anchor=west] {$\mathcal{X}$};
\filldraw[black] (4,1)   node[anchor=west] {$\approx$};
\draw [fill=red!30!,ultra thick] (5,-1) rectangle (5.4,2);
\filldraw[black] (4.9,-1.5)   node[anchor=west] {$\mathbf{a}_1$};
\draw [fill=red!30!,ultra thick] (5.6,1.6) rectangle (8.6,2);
\filldraw[black] (6.8,1)   node[anchor=west] {$\mathbf{b}_1$};
\draw [fill=red!30!,ultra thick] (5, 2.2) -- (5.4, 2.2) -- (6.8,3) -- (6.4,3) -- (5,2.2);
\filldraw[black] (6,2.4)   node[anchor=west] {$\mathbf{c}_1$};
\filldraw[black] (9,1)   node[anchor=west] {$+$};
\filldraw[black] (10,1)   node[anchor=west] {$\hdots$};
\filldraw[black] (11,1)   node[anchor=west] {$+$};

\draw [fill=red!30!,ultra thick] (12,-1) rectangle (12.4,2);
\filldraw[black] (11.9,-1.5)   node[anchor=west] {$\mathbf{a}_R$};
\draw [fill=red!30!,ultra thick] (12.6,1.6) rectangle (15.6,2);
\filldraw[black] (13.8,1)   node[anchor=west] {$\mathbf{b}_R$};
\draw [fill=red!30!,ultra thick] (12, 2.2) -- (12.4, 2.2) -- (13.8,3) -- (13.4,3) -- (12,2.2);
\filldraw[black] (13.2,2.4)   node[anchor=west] {$\mathbf{c}_R$};
\end{tikzpicture}
\caption{CP decomposition of a third order tensor $\mathcal{X} \in \mathbbm{R}^{n\times p \times d}$ from \eqref{CPDecom}.}
\label{CPFIG}
\end{figure}

Let $\mathcal{X} \in \mathbbm{R}^{I\times J \times K}$  be as above the mode 3 tensor that stacks along the third dimension, the spatio-temporal variables, and let $R$ be the rank of $\mathcal{X}$. Then CPD would represent the tensor $\mathcal{X}$ in the following form
\begin{equation} \label{CPDecom}
\mathcal{X} = \sum_{r=1}^{R}{\mathbf{a}_r \circ \mathbf{b}_r \circ \mathbf{c}_r } = [[\mathbf{A},\mathbf{B}, \mathbf{C}]], 
\end{equation}
where $\mathbf{A}=[\mathbf{a}_1 \hdots \mathbf{a}_R] \in \mathbbm{R}^{I \times R}$, $\mathbf{B}=[\mathbf{b}_1 \hdots \mathbf{b}_R] \in \mathbbm{R}^{J \times R}$ and $\mathbf{C}=[\mathbf{c}_1 \hdots \mathbf{c}_R] \in \mathbbm{R}^{K \times R}$ are called factor matrices.
To determine the factor matrices of CPD in \eqref{CPDecom}, the Alternating Least Square (ALS) approach is typically applied. ALS consists of finding each factor matrix in turn through gradient descent while fixing the other two \citep{liu2022tensor}.
The CP decomposition from \eqref{CPDecom} is illustrated in Fig.~\ref{CPFIG}.

While the Alternating Least Squares (ALS) algorithm is widely used for computing CP decomposition due to its simplicity and empirical success, its theoretical foundation, particularly regarding spatio-temporal data remain underdeveloped, especially in noisy, non-orthogonal, and higher-rank settings. It is also known that the loss surface can be highly non-convex with exponentially many local minima \citep{arous2018landscapespikedtensormodel} when the signal to noise ratio is not sufficiently strong, making convergence to the true loading vectors theoretically uncertain with ALS. Initialization is then crucial for the success of ALS. For rank being 1 methods like Higher-Order Singular Value Decomposition (HOSVD) have proven to be an effective initializations; see \cite{8368145}.  This is the motivation for our method were the initialization regularize to all rank. In our method, first we find the Spatio-Temporal Principal Component, then find the factor matrices using the components, and finally using those factor as initial factor to find the CP factor via ALS. Recently, Tang et al. \cite{tang2025revisitcptensordecomposition} proposed a hybrid initialization strategy—Tucker-based Approximation with Simultaneous Diagonalization (TASD). The method uses Tucker decomposition to compress the tensor followed by SimDiag \citep{doi:10.1137/0614071,7869678} applied to the low-dimensional core establishing a statistical guarantees for general rank R. However, currently there are no initialization method that is built specifically to consider the spatial and temporal component of a spatio-temporal data. Our method reaches the same level of relative error like TASD even without the extra step of applying SimDiag which increase the time the initialization of the factor matrices for CP decomposition actually takes. The initialization step with STPCA is much faster compared to TASD.

\subsection{K means clustering}\label{kmeans}

K-means is an unsupervised clustering algorithm, which divides unlabeled data into a certain number (commonly denoted as $K$) of distinct groupings by minimizing Euclidean distances between them \cite{bauckhage2015k}. It assumes that the closer the distance between two points, the greater the similarity. k-means is primarily designed for low-array datasets, making it unsuitable for handling high-dimensional tensor datasets. Applying tensor decomposition model into K-means clustering algorithm can accurately find the multilinear subspaces of data and cluster these high-order data at the same time. In our analysis the goal is to find the subspace of our high-dimensional data using CPD and use k-means clustering to cluster the latent factor along the tensor mode into $k$ groups. K- means can be generalized to tensor spaces using a tensor decomposition. Considering our third order tensor $\mathbbm{R}^{n\times p \times d}$ with CPD factor  $\mathbf{A} \in \mathbbm{R}^{I \times R}$, $\mathbf{B} \in \mathbbm{R}^{J\times R}$, $\mathbf{C} \in \mathbbm{R}^{K \times R}$, and $R$ is the CP rank. Clustering is applied to the rows of $A$.
\begin{equation}
    \min_{\mathbf{A}, \mathbf{B}, \mathbf{C}, \mathbf{S}, \mathbf{C}, \mathbf{D}}
    \|\mathbf{X}_{1} - \mathbf{D} \mathbf{A} (\mathbf{C} \odot \mathbf{B})^T \|^2_F
    + \lambda \|\mathbf{A} - \mathbf{S} \mathbf{C}\|^2_F
    + \eta(\|\mathbf{B}\|^2_F + \|\mathbf{C}\|^2_F).
\end{equation}
The second term represents the \textit{K-means penalty}, while the last term regularizes the factors. This optimization problem is done under the following constraints:
\begin{equation}
    \mathbf{A}, \mathbf{B}, \mathbf{C} \geq 0, \quad \|\mathbf{S}(n, :)\|_0 = 1.
\end{equation}
The algorithm iterates updates for clustering, decomposition, and factor matrices until convergence.
\\~\\
\textbf{Performance Measure.} To evaluate how well the clusters from kmeans are separated for CP Decomposition, we use the silhouette score. The Silhouette coefficient is based on the dissimilarity of a component to all other components within the same cluster (intra-cluster distance) and the distance of the component to the nearest cluster (inter-cluster distance). The intra-cluster distances and the nearest-cluster distances are averaged for each component (denoted as ‘a’ and ‘b’, respectively), and the Silhouette Coefficient is computed by $\frac{b - a}{\max(a, b)}$ \citep{ROUSSEEUW198753}. This assess how well-separated and compact the clusters are. 

\section{Results and discussion}\label{result}

In this section we have two types of numerical experiments for testing the performance of our algorithm. The first one is using our dataset as described in \cref{regdata} to run a CPD decomposition and compare performance using the relative error given by $\frac{\lVert \mathcal{X}_{\text{estimate}} - \mathcal{X} \rVert_{2}}
     {\lVert \mathcal{X} \rVert_{2}}.$ The second experiment is a Kmeans clustering analysis done to again test the performance of our method compared to existing ones.

\subsection{STPCA-CPD using Alternating Least Square method}

In the following experiments we applied the STPCA-CPD algorithm to the real dataset describe in \cref{regdata}. We did the decomposition varying the number of components (i.e. the rank). We also compare the performance of our algorithm with the well known HOSVD+CPD where the initialization of the factor matrices is done using Higher Order Singular Value Decomposition (HOSVD), and a random initialization coupled with ALS to find the CP factor matrices. We compute the reconstruction error using the relative error for each method. All the decompositions for HOSVD+CPD,  STPCA+CPD, and random+CPD were computed using the matlab tensor toolbox \citep{doi:10.1137/07070111X, tensor_toolbox_2024} and the tensorLab toolbox \cite{TensorLab}.

\begin{table}[ht]
\caption{Relative reconstruction errors for CPD with ALS where $\mathcal{X} \in \mathbbm{R}^{13149 \times 1054 \times 3}$.}\label{tab:reconst:errors}
\centering
\begin{tabular}{ |p{2cm}||p{3cm}|p{3cm}|p{3cm}|  }
  \hline
Initialization & Relative error for rank = 2 & Relative error for rank = 3   \\ 
  \hline
 HOSVD &  0.4928  & 0.3832\\ 
  Random & 0.4930 & 0.3849 \\ 
  STPCA & 0.4910  & 0.3816\\
   \hline
\end{tabular}

\end{table}

The overall the CPD with STPCA as initializer perform better with a relative error lower than the other methods see Table \ref{tab:reconst:errors}. Note that for this analysis we only used 2 and 3 components but still ended up with a good results. The method is applicable to $n$ number of components.

\subsection{Clustering analysis of the CP factors per mode}

To continue this analysis we are now looking at the K-means clustering (see \cref{kmeans}) of the CPD factors with the initialization methods used above. The goal in this case is to use the components from all the method in mode 1 (temporal mode) and mode 2 (spatial mode), and do a K-means clustering analysis. We use the silhouette score as described in \cref{kmeans}. The k-means analysis run through a range of k from 2 to 12 for all the decompositions. Table \ref{tab:kmeans_2} shows the results for rank 2 and Table \ref{tab:kmeans_3} shows the results for rank 3. In both cases, STPCA+CPD outperform the other methods.

\begin{table}[ht]
\caption{Silhouette scores for k-means clustering applied to CPD components with rank 2 }\label{tab:kmeans_2}
\centering
\begin{tabular}{|p{2cm}|p{3cm}|p{1cm}|p{3cm}|p{1cm}|}
  \hline
  Initialization & Silhouette Scores for mode 1 & Best k & Silhouette Scores for mode 2  & Best k \\
  \hline
  HOSVD & 0.6484 & 2 & 0.5872 & 2 \\
  Random & 0.6579 & 2 & 0.6002 & 2 \\
  STPCA &  0.7990 & 2 & 0.6184 &4\\
  \hline
\end{tabular}

\end{table}

\begin{table}[ht]
\caption{Silhouette scores for k-means clustering applied to CPD components with rank 3 }\label{tab:kmeans_3}
\centering
\begin{tabular}{|p{2cm}|p{3cm}|p{1cm}|p{3cm}|p{1cm}|}
  \hline
  Initialization & Silhouette Scores for mode 1 & Best k & Silhouette Scores for mode 2  & Best k \\
  \hline
  HOSVD & 0.4539 & 3 & 0.6422 & 2 \\
  Random & 0.4968	 & 3 & 0.6566 & 2 \\
  STPCA &0.5795  & 2 &0.6589  &2\\
  \hline
\end{tabular}

\end{table}

\section{Conclusions}

In this study, we employed a Spatio-Temporal Principal Component Analysis (STPCA) to initialize the factors in our CANDECOMP/PARAFAC (CP) decomposition for better result using the Alternating Least Squares (ALS) algorithm. The numerical experiments yielded promising results, demonstrating a distinct advantage in leveraging the spatio-temporal dependencies inherent in the data into the CP factorization process. Model performance was assessed through the computation of the relative reconstruction error for each configuration.
Furthermore, we applied k-means clustering to the extracted factor matrices and evaluated clustering quality using silhouette scores. The proposed STPCA+CPD approach consistently outperformed baseline methods, confirming the benefits of our initialization strategy.
For future work, this framework could be extended by integrating deep learning architectures to enhance its capacity to capture complex spatio-temporal dynamics. Additionally, we plan to investigate more robust spatio-temporal tensor decomposition schemes and to generalize the tensor framework for forecasting and downscaling applications.

\bibliography{sn-bibliography}

@article{ROUSSEEUW198753,
title = {Silhouettes: A graphical aid to the interpretation and validation of cluster analysis},
journal = {Journal of Computational and Applied Mathematics},
volume = {20},
pages = {53-65},
year = {1987},
issn = {0377-0427},
doi = {https://doi.org/10.1016/0377-0427(87)90125-7},
url = {https://www.sciencedirect.com/science/article/pii/0377042787901257},
author = {Peter J. Rousseeuw}
}

@article{bauckhage2015k,
  title={K-means clustering is matrix factorization},
  author={Bauckhage, Christian},
  journal={arXiv preprint arXiv:1512.07548},
  year={2015}
}

@article{Lathauwer1996FromMT,
author = {Lathauwer, Lieven and De Moor, Bart},
year = {1997},
month = {01},
pages = {1-15},
title = {From matrix to tensor: Multilinear algebra and signal processing},
journal = {Mathematics in Signal Processing IV}
}

@article{article12,
author = {Scott C. Sheridan and Cameron C. Lee},
title ={The self-organizing map in synoptic climatological research},
journal = {Progress in Physical Geography: Earth and Environment},
volume = {35},
number = {1},
pages = {109-119},
year = {2011},
doi = {10.1177/0309133310397582}
}

@book{Aggarwal2015,
  title     = {Data Mining: The Textbook},
  author    = {Charu C. Aggarwal},
  year      = {2015},
  publisher = {Springer},
  address   = {Cham},
  isbn      = {978-3-319-14142-8},
  url       = {https://doi.org/10.1007/978-3-319-14142-8}
}

@book{liu2022tensor,
  title={Tensor computation for data analysis},
  author={Liu, Y. and Liu, J. and Long, Z. and Zhu, C.},
  year={2022},
  publisher={Springer},
  address = {Cham},
URL = {https://doi.org/10.1007/978-3-030-74386-4}
}

@article{cliff1ord,
  title={The Problem of Spatial Autocorrelation},
  author={CLIFF, ADETORD},
  journal={London Papers in Regional Science, Pion, London},
  pages={25--55}
}

@article{chen2023spatial,
  title={Spatial autocorrelation equation based on Moran’s index},
  author={Chen, Yanguang},
  journal={Scientific reports},
  volume={13},
  number={1},
  pages={19296},
  year={2023},
  publisher={Nature Publishing Group UK London}
}

@misc{tian2023tensor,
      title={Tensor BM-Decomposition for Compression and Analysis of Spatio-Temporal Third-order Data}, 
      author={Fan Tian and Misha E. Kilmer and Eric Miller and Abani Patra},
      year={2023},
      eprint={2306.09201},
      archivePrefix={arXiv},
      primaryClass={math.NA}
}

@article{article09,
author = {Gong, Xiaofeng and Richman, Michael},
year = {1995},
month = {03},
pages = {897-931},
title = {On the Application of Cluster Analysis to Growing Season Precipitation Data in North America East of the Rockies},
volume = {8},
journal = {Journal of Climate},
doi = {10.1175/1520-0442(1995)008<0897:OTAOCA>2.0.CO;2}
}

@misc{climate,
      title={Impacts, Risks, and Adaptation in the United States: Fourth National Climate Assessment, Volume II},
      author={D.R Reidmiller and C.W. Avery and D.R Easterling and K.E. Kunkel and K.L.M. Lewis and T.K. Maycock and B.C. Stewart. and U.S. Global Change Research Program and National Oceanic and Atmospheric Administration and National Aeronuatics and Space Administration},  
      year={2017},
      URL = { https://repository.library.noaa.gov/view/noaa/19487}}

@misc{article,
      title={Gridspec: A standard for the description of grids used in Earth System models}, 
      author={Venkatramani Balaji and Alistair Adcroft and Zhi Liang},
      year={2019},
      eprint={1911.08638},
      archivePrefix={arXiv},
      primaryClass={physics.ao-ph},
      url={https://arxiv.org/abs/1911.08638}, 
}

@INPROCEEDINGS{7846858,
  author={Virta, Joni and Taskinen, Sara and Nordhausen, Klaus},
  booktitle={2016 IEEE Signal Processing in Medicine and Biology Symposium (SPMB)}, 
  title={Applying fully tensorial ICA to fMRI data}, 
  year={2016},
  volume={},
  number={},
  pages={1-6},
  doi={10.1109/SPMB.2016.7846858}}

@article{Cartone03042021,
author = {Alfredo Cartone and Paolo Postiglione},
title = {Principal component analysis for geographical data: the role of spatial effects in the definition of composite indicators},
journal = {Spatial Economic Analysis},
volume = {16},
number = {2},
pages = {126--147},
year = {2021},
publisher = {RSA Website},
doi = {10.1080/17421772.2020.1775876}

}

@article{Harsh,
  title={Foundations of the PARAFAC procedure: Models and conditions for an" explanatory" multimodal factor analysis},
  author={Harshman, R.A.},
  journal={UCLA Working Papers in Phonetics},
  volume={16},
  number={1},
  pages={84},
  year={1970}
}

@article{Kol,
author = {Tian, Fan and Kilmer, Misha E. and Miller, Eric and Patra, Abani},
title = {Tensor BM-Decomposition for Compression and Analysis of Video Data},
journal = {Numerical Linear Algebra with Applications},
volume = {32},
number = {6},
pages = {e70043},
doi = {https://doi.org/10.1002/nla.70043},
year = {2025}
}

@ARTICLE {Xu,
author = {J. Xu and J. Zhou and P. Tan and X. Liu and L. Luo},
journal = {IEEE Transactions on Knowledge and Data Engineering},
title = {Spatio-Temporal Multi-Task Learning via Tensor Decomposition},
year = {2021},
volume = {33},
number = {06},
issn = {1558-2191},
pages = {2764-2775},
doi = {10.1109/TKDE.2019.2956713},
publisher = {IEEE Computer Society},
address = {Los Alamitos, CA, USA},
month = {jun}
}

@article{doi:10.1137/07070111X,
author = {Kolda, Tamara G. and Bader, Brett W.},
title = {Tensor Decompositions and Applications},
journal = {SIAM Review},
volume = {51},
number = {3},
pages = {455-500},
year = {2009},
doi = {10.1137/07070111X}
}

@article{Hitchcock1927TheEO,
  title={The Expression of a Tensor or a Polyadic as a Sum of Products},
  author={Frank Lauren Hitchcock},
  journal={Journal of Mathematics and Physics},
  year={1927},
  volume={6},
  pages={164-189},
  url={https://api.semanticscholar.org/CorpusID:124183279}
}

@article{Carroll,
  title={Analysis of individual differences in multidimensional scaling via an N-way generalization of “Eckart-Young” decomposition},
  author={Carroll, J. D. and Chang, J.J.},
  journal={Psychometrika},
  volume={35},
  number={3},
  pages={283--319},
  year={1970},
  publisher={Springer}
}

@article{WangEtal2021,
author = {Wang, Fang and Tian, Di and Lowe, Lisa and Kalin, Latif and Lehrter, John},
title = {Deep Learning for Daily Precipitation and Temperature Downscaling},
journal = {Water Resources Research},
volume = {57},
number = {4},
pages = {e2020WR029308},
doi = {https://doi.org/10.1029/2020WR029308},
year = {2021}
}

@ARTICLE{chen2023fengwu,
       author = {{Chen}, Kang and {Han}, Tao and {Ling}, Fenghua and {Gong}, Junchao and {Bai}, Lei and {Wang}, Xinyu and {Luo}, Jing-Jia and {Fei}, Ben and {Zhang}, Wenlong and {Chen}, Xi and {Ma}, Leiming and {Zhang}, Tianning and {Su}, Rui and {Ci}, Yuanzheng and {Li}, Bin and {Yang}, Xiaokang and {Ouyang}, Wanli},
        title = "{The operational medium-range deterministic weather forecasting can be extended beyond a 10-day lead time}",
      journal = {Communications Earth and Environment},
         year = 2025,
        month = jul,
       volume = {6},
       number = {1},
          eid = {518},
        pages = {518},
          doi = {10.1038/s43247-025-02502-y},
 primaryClass = {cs.AI},
       adsurl = {https://ui.adsabs.harvard.edu/abs/2025ComEE...6..518C}
}

@article{lam2022graphcast,
author = {Lam, R.  and Sanchez-Gonzalez, A.  and Willson, M.  and others },
title = {Learning skillful medium-range global weather forecasting},
journal = {Science},
volume = {382},
number = {6677},
pages = {1416-1421},
year = {2023},
doi = {10.1126/science.adi2336}}

@article{euro,
    author = {Ph. Ciais and M. Reichstein and N. Viovy and A. Granier and J. Ogée and V. Allard and M. Aubinet and N. Buchmann and Chr. Bernhofer and A. Carrara and F. Chevallier and N. De Noblet and A. D. Friend and P. Friedlingstein and T. Grünwald and B. Heinesch and P. Keronen and A. Knohl and G. Krinner and D. Loustau and G. Manca and G. Matteucci and F. Miglietta and J. M. Ourcival and D. Papale and K. Pilegaard and S. Rambal and G. Seufert and J. F. Soussana and M. J. Sanz and E. D. Schulze and T. Vesala and R. Valentini },
    title = {Europe-wide reduction in primary productivity caused by the heat and drought in 2003.},
    journal = {Nature},
    volume = {437},
    pages = {529–533},
    year = {2005}
}

@article{HohleinEtal2020,
author = {Höhlein, K. and Kern, M. and Hewson, T. and Westermann, R.},
title = {A comparative study of convolutional neural network models for wind field downscaling},
journal = {Meteorological Applications},
volume = {27},
number = {6},
pages = {e1961},
doi = {https://doi.org/10.1002/met.1961},
year = {2020}
}

@article{AnnauEtal2023,
  title={Algorithmic hallucinations of near-surface winds: Statistical downscaling with generative adversarial networks to convection-permitting scales},
  author={Annau, Nicolaas J and Cannon, Alex J and Monahan, Adam H},
  journal={Artificial Intelligence for the Earth Systems},
  volume={2},
  number={4},
  pages={e230015},
  year={2023},
  publisher={American Meteorological Society}
}

@book{Goodfellow-et-al-2016,
    title={Deep Learning},
    author={Ian Goodfellow and Yoshua Bengio and Aaron Courville},
    publisher={MIT Press},
    year={2016},
    address   = {Cambridge, MA},
    note={\url{http://www.deeplearningbook.org}},
}

@article{jolliffe2016principal,
    author = {Jolliffe, Ian T. and Cadima, Jorge},
    title = {Principal component analysis: a review and recent developments},
    journal = {Philosophical Transactions of the Royal Society A: Mathematical, Physical and Engineering Sciences},
    volume = {374},
    number = {2065},
    pages = {20150202},
    year = {2016},
    month = {04},
    issn = {1364-503X},
    doi = {10.1098/rsta.2015.0202}
}

@article{Ye_2025, title={A comprehensive review of Principal Component Analysis}, volume={17}, url={https://drpress.org/ojs/index.php/ajst/article/view/32396}, DOI={10.54097/5mmrkr11}, abstractNote={
PCA (Principal Component Analysis) is a method aiming to reduce the dimensions among data analysis, with various applications in neurosciences, finance, and beyond. Data normalization, covariance matrix decomposition, eigenvalue-driven component selection, and other mathematical underpinnings of PCA will be methodically covered in this article. A comparison with SVD decomposition will also be made due to the similarities between the two methods. Additionally, we will discuss contemporary developments like sparse PCA, kernel PCA, and robust PCA that tackle nonlinearity and sparsity by integrating trends like PCA’s integration with deep learning, the variation in applied circumstances, and its use in high-dimensional data presentation. Furthermore, this review will also highlight the inherent limits, such as nonlinearity issues, massive datasets, and data contamination. Throughout investigation, this review serves as a map for the researchers tackling with increasingly complex data environments requiring dimensionality reduction and are not certain with the specific PCA type selected to apply.
}, number={1}, journal={Academic Journal of Science and Technology}, author={Ye, Yubo}, year={2025}, month={Nov.}, pages={224–227} }

@article{moran1950test,
 ISSN = {00063444, 14643510},
 URL = {http://www.jstor.org/stable/2332162},
 author = {P. A. P. Moran},
 journal = {Biometrika},
 number = {1/2},
 pages = {178--181},
 publisher = {[Oxford University Press, Biometrika Trust]},
 title = {A Test for the Serial Independence of Residuals},
 urldate = {2026-03-17},
 volume = {37},
 year = {1950}
}

@book{bishop2006pattern,
  title={Pattern recognition and machine learning},
  author={Bishop, Christopher M and Nasrabadi, Nasser M},
  volume={4},
  number={4},
  year={2006},
  Series= {Information Science and Statistics},
  publisher={Springer New York, NY},
  address   = {New York}
}

@article{gorecki2018selected,
  title={Selected statistical methods of data analysis for multivariate functional data},
  author={G{\'o}recki, Tomasz and Krzy{\'s}ko, Miros{\l}aw and Waszak, {\L}ukasz and Wo{\l}y{\'n}ski, Waldemar},
  journal={Statistical Papers},
  volume={59},
  number={1},
  pages={153--182},
  year={2018},
  publisher={Springer}
}

@article{HannachiEtal2007,
author = {Hannachi, A. and Jolliffe, I. T. and Stephenson, D. B.},
title = {Empirical orthogonal functions and related techniques in atmospheric science: A review},
journal = {International Journal of Climatology},
volume = {27},
number = {9},
pages = {1119-1152},
doi = {https://doi.org/10.1002/joc.1499},
year = {2007}
}

@article{gorecki2012functional,
  title={Functional principal components analysis},
  author={G{\'o}recki, Tomasz and Krzy{\'s}ko, Miros{\l}aw},
  journal={Data analysis methods and its applications},
  pages={71--87},
  year={2012},
  publisher={Beck, Warsaw}
}

@article{Krzyko02012024,
author = {Mirosław Krzyśko and Peter Nijkamp and Waldemar Ratajczak and Waldemar Wołyński and Beata Wenerska},
title = {Spatio-temporal principal component analysis},
journal = {Spatial Economic Analysis},
volume = {19},
number = {1},
pages = {8--29},
year = {2024},
publisher = {Routledge},
doi = {10.1080/17421772.2023.2237532}

}

@article {MonahanEtal2009,
      author = "A. H. Monahan and J. C. Fyfe and M. H. P. Ambaum and D. B. Stephenson and G. R. North",
      title = "Empirical Orthogonal Functions: The Medium is the Message",
      journal = "Journal of Climate",
      year = "2009",
      publisher = "American Meteorological Society",
      address = "Boston MA, USA",
      volume = "22",
      number = "24",
      doi = "10.1175/2009JCLI3062.1",
      pages=      "6501 - 6514",
      url = "https://journals.ametsoc.org/view/journals/clim/22/24/2009jcli3062.1.xml"
}

@article{hotelling1933analysis,
  title={Analysis of a complex of statistical variables into principal components.},
  author={Hotelling, Harold},
  journal={Journal of educational psychology},
  volume={24},
  number={6},
  pages={417},
  year={1933},
  publisher={Warwick \& York}
}

@ARTICLE{8368145,
  author={Zhang, Anru and Xia, Dong},
  journal={IEEE Transactions on Information Theory}, 
  title={Tensor SVD: Statistical and Computational Limits}, 
  year={2018},
  volume={64},
  number={11},
  pages={7311-7338},
  doi={10.1109/TIT.2018.2841377}}

@article{arous2018landscapespikedtensormodel,
  title={The landscape of the spiked tensor model},
  author={Arous, Gerard Ben and Mei, Song and Montanari, Andrea and Nica, Mihai},
  journal={Communications on Pure and Applied Mathematics},
  volume={72},
  number={11},
  pages={2282--2330},
  year={2019},
  publisher={Wiley Online Library}
}

@article{doi:10.1137/0614071,
author = {Leurgans, S. E. and Ross, R. T. and Abel, R. B.},
title = {A Decomposition for Three-Way Arrays},
journal = {SIAM Journal on Matrix Analysis and Applications},
volume = {14},
number = {4},
pages = {1064-1083},
year = {1993},
doi = {10.1137/0614071}
}

@misc{tensor_toolbox_2024,
  author       = {Brett W. Bader and Tamara G. Kolda and others},
  title        = {{Tensor Toolbox for MATLAB, Version 3.8}},
  howpublished = {\url{https://www.tensortoolbox.org/}},
  year         = {2024}
}

@INPROCEEDINGS{7869678,
  author={Naskovska, Kristina and Haardt, Martin},
  booktitle={2016 50th Asilomar Conference on Signals, Systems and Computers}, 
  title={Extension of the semi-algebraic framework for approximate CP decompositions via simultaneous matrix diagonalization to the efficient calculation of coupled CP decompositions}, 
  year={2016},
  volume={},
  number={},
  pages={1728-1732},
  doi={10.1109/ACSSC.2016.7869678}}

@misc{tang2025revisitcptensordecomposition,
      title={Revisit CP Tensor Decomposition: Statistical Optimality and Fast Convergence}, 
      author={Runshi Tang and Julien Chhor and Olga Klopp and Anru R. Zhang},
      year={2025},
      eprint={2505.23046},
      archivePrefix={arXiv},
      primaryClass={stat.ME},
      url={https://arxiv.org/abs/2505.23046}, 
}

@article{nacordex,
  author = {Mearns, L. and McGinnis, S. and Korytina, D. and others },
  title = {The NA-CORDEX dataset, version 1.0},
  year = {2017},
  doi = {https://doi.org/10.5065/D6SJ1JCH},
journal = {NCAR Climate Data Gateway Boulder CO, accessed 08/11/2021}}

@article{KRUSKAL197795,
title = {Three-way arrays: rank and uniqueness of trilinear decompositions, with application to arithmetic complexity and statistics},
journal = {Linear Algebra and its Applications},
volume = {18},
number = {2},
pages = {95-138},
year = {1977},
issn = {0024-3795},
doi = {https://doi.org/10.1016/0024-3795(77)90069-6},
url = {https://www.sciencedirect.com/science/article/pii/0024379577900696},
author = {Joseph B. Kruskal}
}

@inproceedings{TensorLab,
author = {Vervliet, Nico and Debals, Otto and Lathauwer, Lieven},
year = {2016},
month = {11},
pages = {1733-1738},
title = {Tensorlab 3.0 — Numerical optimization strategies for large-scale constrained and coupled matrix/tensor factorization},
doi = {10.1109/ACSSC.2016.7869679}
}

\backmatter


\subsection*{Acknowledgements}The author would like to thank Professor Julie Carreau for helpful discussion concerning this project.
\end{document}